\documentclass[12pt, a4paper]{article}

\usepackage{graphicx}
\usepackage{amsmath}
\usepackage{geometry}
\usepackage[english,activeacute]{babel}
\usepackage{siunitx}
\usepackage{pgfplots}
\usepackage[bottom]{footmisc}
\usepackage{float}
\usepackage{fancyhdr}
\usepackage{setspace}
\usepackage{rotating}
\usepackage{epstopdf} 
\usepackage{tgadventor}
\usepackage{epsfig}
\usepackage{amsfonts}
\usepackage{mathrsfs}
\usepackage{hyperref}
\usepackage[affil-it]{authblk}
\usepackage{xcolor}
\usepackage{empheq}
\usepackage{mathtools}
\usepackage[nottoc]{tocbibind}
\usepackage{verbatim}
\usepackage{fancyref}
\usepackage{subcaption}
\usepackage{bigints}
\usepackage[utf8]{inputenc}
\usepackage{minibox}
\usepackage{titlesec}
\usepackage{tikz,pgfplots}
\usepackage[makeroom]{cancel}
\usepackage{epsfig}
\usepackage{graphicx}
\usepackage[numbers,sort&compress]{natbib}
\usepackage{color}
\usepackage{tcolorbox}
\usepackage{csquotes}
\usetikzlibrary{snakes}

\usepackage{dsfont}
\usepackage{relsize}

\usetikzlibrary{fillbetween,backgrounds}

\usepackage{leftidx}

\usepgfplotslibrary{fillbetween}
\usetikzlibrary{patterns}

\usepackage{float}
\usepackage[font=small,labelfont=bf]{caption}
\usetikzlibrary{arrows,shapes}

\usepackage{braket}
\hypersetup{colorlinks ,urlcolor=blue,anchorcolor=blue ,citecolor=blue!70!black ,filecolor=blue ,linkcolor=blue!70!black ,menucolor=blue ,pagecolor=blue ,linktocpage=true ,pdfproducer=medialab ,pdfa=true
}

\DeclareFixedFont\trfont{OT1}{phv}{b}{sc}{11}

\newcommand{\RomanNumeralCaps}[1]
{\MakeUppercase{\romannumeral #1}}

\definecolor{atomictangerine}{rgb}{1.0, 0.6, 0.4}
\definecolor{asparagus}{rgb}{0.53, 0.66, 0.42}
\definecolor{bisque}{rgb}{1.0, 0.89, 0.77}
\definecolor{mint}{rgb}{0.24, 0.71, 0.54}

\usepackage{amsmath}
\usepackage{amsfonts}
\usepackage{amssymb}
\usepackage{graphicx, rotating}
\usepackage{epstopdf}
\usepackage{epsfig}
\usepackage{latexsym}
\usepackage{graphicx}
\usepackage{color}
\usepackage{amsmath,bm,amssymb}
\usepackage{soul}
\usepackage{slashed}
\usepackage{float}
\usepackage{hyperref}
\hypersetup{colorlinks, citecolor=bluscuro, linkcolor=black, urlcolor=bluscuro}
\definecolor{rossos}{cmyk}{0,1,1,0.55}
\definecolor{bluscuro}{rgb}{0.15, 0.2, .85}
\definecolor{bluchiaro}{cmyk}{1,.3,0.,0.1}
\definecolor{brown}{rgb}{0.6, 0.14, 0.14}

\definecolor{paleaqua}{rgb}{0.74, 0.83, 0.9}

\newcommand{\be}{\begin{equation}}
\newcommand{\ee}{\end{equation}}
\newcommand{\bea}{\begin{eqnarray}}
\newcommand{\eea}{\end{eqnarray}}

\begin{document}

\begin{titlepage}
\begin{flushright}
IFT-UAM/CSIC-19-83
\end{flushright}
\vspace{.3in}

\vspace{1cm}
	\begin{center}
	{\Large\bf\color{black} Holographic Bulk Reconstruction And Cosmological Singularities}\\
	
	\bigskip\color{black}
	\vspace{1cm}{
		{\large Jos\'e ~L.~F. Barb\'on and  Martin Sasieta}
		\vspace{0.3cm}
	} \\[7mm]
	{\it {Instituto de F\'{\i}sica Te\'orica,  IFT-UAM/CSIC}}\\
	{\it {C/ Nicol\'as Cabrera 13, Universidad Aut\'onoma de Madrid, 28049 Madrid, Spain}}\\
	{\it E-mail:} \href{mailto:jose.barbon@uam.es}{\nolinkurl{jose.barbon@csic.es}}, \href{mailto:martin.sasieta@uam.es}{\nolinkurl{martin.sasieta@uam.es}}\\

\end{center}
\bigskip

\vspace{.4cm}

	\begin{abstract}
	We study the structure of entanglement wedges in the Kasner-AdS geometry, which provides an example of AdS/CFT engineered cosmological singularity. We investigate the specific limitations of causal reconstruction methods, imposed by the presence of the cosmological singularities, and we show the supremacy of modular reconstruction. This model provides an example where  modular reconstruction based on a proper operator subalgebra is more powerful than the strongest possible causal reconstruction, based on the complete operator algebra.   
	\end{abstract}

\bigskip

\end{titlepage}


\section{Introduction}
\noindent

A significant intuitive clue into the decoding of the holographic AdS/CFT dictionary has come into gradual focus in recent years. A number of concrete constructions of approximately local physics in the bulk have been proposed, using states and operator algebras in the CFT. 

Perhaps the most explicit  of such constructions is the so called HKLL prescription (cf. \cite{Banks:1998dd,Bena:1999jv,Hamilton:2005ju,Hamilton:2006az,Kabat:2011rz,Heemskerk:2012mn,Kabat:2012av,Kabat:2012hp,Kabat:2013wga}) which  builds  local  bulk operators in a $1/N$ expansion in terms of a CFT operator algebra associated to causal domains of the CFT spacetime. This method uses the machinery of Green's functions in the bulk background to produce expressions of the form
\begin{equation}\label{hkll}
\phi(X) \big|_{CW_R} = \int_{D_R} f(X; x) \,{\cal O}(x) = \int dt \int_{R} dx_R \,f(X; x_R, t) \;{\cal O}(x_R, t)\;,
\end{equation}
where $D_R$ is the causal development of a spacelike region $R$ on the CFT spacetime and $CW_R$ is the causal wedge in the bulk of this set, that is to say, the intersection of past and future of $D_R$ in the bulk: $J^+ (D_R) \cap J^- (D_R)$. The formula \eqref{hkll} constructs a local field on $CW_R$ (in the sense of low-energy effective theory) out of CFT operators ${\cal O}$ defined on $D_R$, using   an appropriate Green's function $f$ and a suitable amount of gauge dressing that we are suppressing in the discussion. The second expression in \eqref{hkll} makes explicit that the set of local operators on $D_R$ can be obtained by Heisenberg time-flow from the set of local operators on $R$, once we are given a time foliation of $D_R$ which contains $R$ itself.

The CFT operator algebra on $R$ is known to be more powerful than what is revealed by the expression \eqref{hkll}. In fact, it has been argued that reconstruction of local fields beyond $CW_R$ is possible by replacing the standard Hamiltonian evolution of the CFT by the modular evolution, which uses the so-called modular Hamiltonian 
$$
K_R = -{1\over 2\pi} \log \,\rho_R\;,
$$
derived from the density matrix on region $R$. The modular Hamiltonian $K_R$ is in general non-local, but the modular analog of Heisenberg operators 
$$
{\cal O}_R (s) = e^{isK_R} \,{\cal O}_R \, e^{-isK_R}\;,
$$
can be used to write down a generalization of \eqref{hkll} of the form
\begin{equation}\label{modular}
\phi(X) \big|_{EW_R} =\int ds  \int_{R} dx_R\, g (X; x_R, s) \,{\cal O}(x_R, s)\;.
\end{equation}
In this case, the smearing function $g(X; x_R, s)$ can be obtained from the powerful statement of \cite{Jafferis:2015del} that modular flow restricted to appropriate low-energy operators is equal to its bulk counterpart defined on the `entanglement wedge' (cf. also \cite{Jafferis:2014lza,Dong:2016eik}). Even more ambitiously, it has been argued that a complete reconstruction is possible wherever the bulk low-energy effective field theory is defined, provided one
uses sufficiently state-dependent prescriptions (cf. \cite{Papadodimas:2012aq,Papadodimas:2013jku,deBoer:2018ibj,deBoer:2019kyr}). This entanglement wedge,   $EW_R$,  is defined as the bulk causal domain of the spacelike region $r$ whose boundary has components $\partial r = R \cup {\chi_R}$, where $\chi_R$ is the  HRT entangling surface anchored on $R$. 
In the leading large-$N$ approximation, in which bulk fields are free, one can give fairly explicit constructions of the kernel $g(X, x_R, s)$ using the gaussian character of the bulk quantum states \cite{Faulkner:2017vdd}.

In this article we investigate effects of cosmological singularities on these reconstruction methods. By cosmological singularity in the holographic context we mean one that is visible over all energy scales of the CFT, in particular in the UV (or boundary) description. Such singular states are typically engineered by a singular driving of the CFT by a time-dependent operator. 
In this note we examine a particular example of AdS/CFT-engineered cosmological singularity based on a marginal operator driving.   On the QFT side we consider a holographic CFT on a Kasner metric, which is Ricci flat, homogeneous,  but otherwise has a generic spacelike singularity. The bulk background is an AdS cone over the Kasner metric, known as the Kasner-AdS background which extends the boundary singularity into the bulk (cf. \cite{Engelhardt:2013jda,Engelhardt:2014mea,Engelhardt:2015gta,Barbon:2015ria} for previous holographic studies of this background). 

We will show that the Kasner-AdS model illustrates an interesting  phenomenon: the presence of the singularity  puts limits to the power of causal reconstructions, even if we allow ourselves to use the full CFT operator algebra on an infinite Cauchy slice.  On the other hand, it is found that modular reconstruction in this background, while based on a {\it finite-region}  operator subalgebra,  is still capable of going beyond the most powerful causal reconstruction. This is a somewhat stronger statement than the usual fact that the entanglement wedge contains the causal wedge, $CW_R \subset EW_R$, for the same $R$ (cf. \cite{Hubeny:2012wa,Wall:2012uf,Headrick:2014cta}).

\section{The Kasner-AdS state}
\noindent

\begin{figure}[h]
	\centering
	\begin{tikzpicture}

	
	\draw [line width = .3mm, gray,dashed] plot [smooth, tension=1] (0,7)..controls (-.75,5.5) and (-1.4,1)..(-1.5,0);
	\draw [line width = .3mm, gray,dashed] plot [smooth, tension=1] (0,7)..controls (-1.5,5)and (-2.9,1)..(-3,0);
	\draw [line width = .3mm, gray,dashed] plot [smooth, tension=1] (0,7)..controls (-2.25,4.5)and (-4.4,1)..(-4.5,0);
	\draw [line width = .3mm, gray,dashed] plot [smooth, tension=1] (0,7)..controls (-3,4) and (-5.9,1)..(-6,0);
	
	\draw [line width = .3mm, gray,dashed] plot [smooth, tension=1] (-7,0)..controls (-5.5,.75) and (-1,1.4)..(0,1.5);
	\draw [line width = .3mm, gray,dashed] plot [smooth, tension=1] (-7,0)..controls (-5,1.5)and (-1,2.9)..(0,3);
	\draw [line width = .3mm, gray,dashed] plot [smooth, tension=1] (-7,0)..controls (-4.5,2.25)and (-1,4.4)..(0,4.5);
	\draw [line width = .3mm, gray,dashed] plot [smooth, tension=1] (-7,0)..controls (-4,3) and (-1,5.9)..(0,6);
	
	\draw[line width = .7mm,red!90!black] (-7,0)   -- (0,7);
	\draw[line width = .8mm, snake = zigzag,red!90!black]    (0,0) -- (-7,0); 
	
	\draw[line width = .5mm] (0,7) -- (0,0);
	
	
	\node[rotate = 45] at (-3.7,3.7) {$t=+\infty$};
	\node at (-3.5,-.4) {$t=0$};
	\node[rotate = 90] at (.3,3.2) {$z=0$};

	\end{tikzpicture}
	\caption{Penrose diagram of the $(t,z)$-projection of Kasner-AdS for $t>0$. Horizontal (vertical) dashed lines are constant $t$ (constant $z$). The diagram for $t<0$ is just the $t\rightarrow -t$ reflection.}
	\label{Penrosediag}
\end{figure}
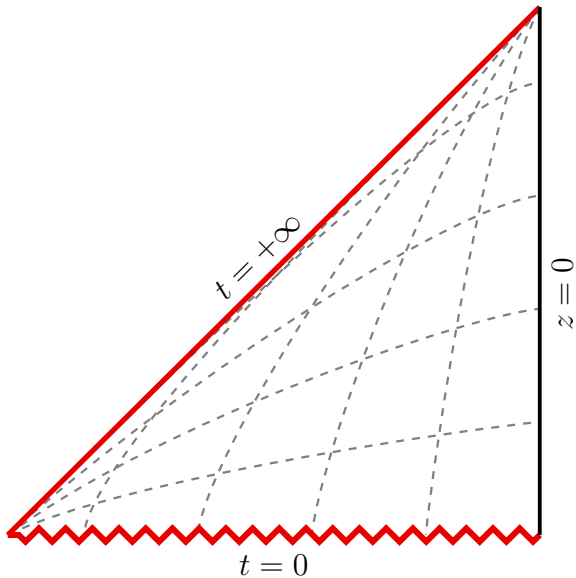

Let us consider a holographic CFT defined on the Kasner spacetime 
\begin{equation}\label{kasner}
ds^2_{\rm CFT} = -dt^2 + \sum_{j=1}^{d-1} t^{\,2p_j} \; dx_j^2\;, 
\end{equation}
where we have chosen to measure time in units of the average expansion rate. Despite the fact that the CFT spacetime has no dynamical gravity, we shall choose the constants $p_j$ satisfying the standard sum rules $\sum_j \,p_j = \sum_j \,p_j^2 =1$, so that the metric is spatially homogeneous, Ricci flat and singular at $t=0$. At least one $p_j$ is negative, which implies one direction of contraction out of the bang towards the future, or one direction of expansion into the crunch from the past.    For any such Ricci flat boundary metric, we can immediately manufacture a bulk solution with constant (negative) bulk curvature in $d+1$ dimensions: 
\begin{equation}\label{bulkone}
ds^2 = {dz^2 + ds_{\rm CFT}^2 \over z^2}
\;, \end{equation}
where we measure the radial coordinate $z>0$ in units of the radius of curvature. A Penrose diagram showing the ${\bf R}^{d-2}$-invariant causal structure is shown in Figure \ref{Penrosediag}.

In this spacetime, the $t=0$ singularity of the CFT metric  penetrates into the full $t=0$ slice of the bulk spacetime. In addition to this $t=0$ singularity, there are milder tidal singularities at $t=\pm \infty$.  At any rate, this background defines a certain holographic state on the CFT side and an associated `code subspace' where low-energy bulk operators may be defined implicitly by the standard machinery of low-energy effective QFT in curved spacetime.

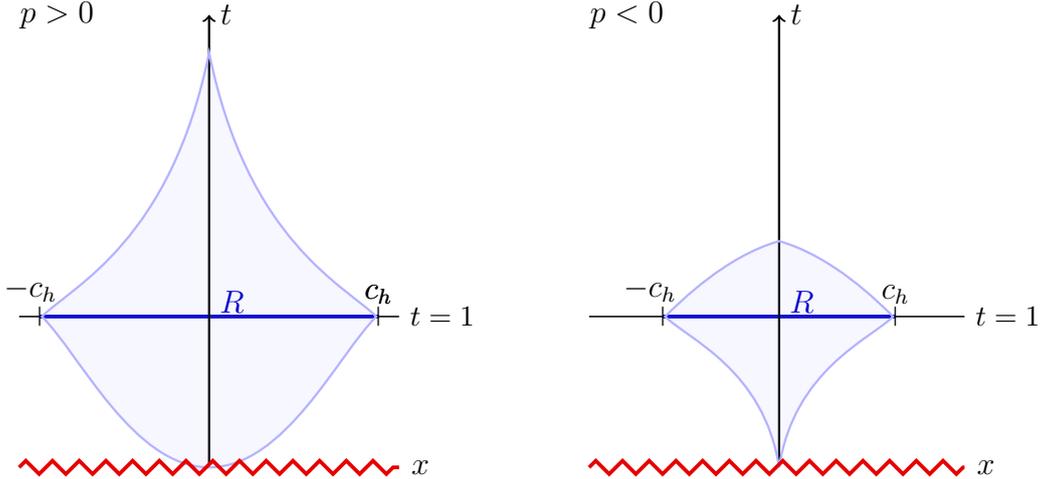
\begin{figure}[h]
	\centering
	\begin{tikzpicture}
	\begin{scope}[xshift = -3.75cm]
	\draw[line width = .2mm] (-2.5,2) -- (2.5,2) node[right]{\small $t=1$};
	\draw[line width = .5mm,blue!80!black] (-2.23,2) -- (2.23,2);
		\draw[->,line width = .3mm] (0,0.03) -- (0,6) node[right]{ $t$};
		\draw [line width = .3mm,blue!30!white, fill,fill opacity = 0.1] (0,0)..controls(-1,0) and (-1.7,1.5).. (-2.2,2) ..controls(-1.8,2.4) and (-.5,3).. (0,5.5)..controls(.5,3) and (1.8,2.4)..(2.2,2)..controls(1.7,1.5) and (1,0).. (0,0);
	   \draw[line width = .5mm,snake = zigzag,red!90!black] (-2.5,0) -- (2.5,0)  node[right] {$\color{black}{x}$};
	   	\node[rotate = 90] at (-2.23,2) {$-$};
	   	\node[rotate = 90] at (2.23,2) {$-$};
	   	 \node at (-2.35,2.35) {$-c_h$};
	   	\node at (2.23,2.3) {$c_h$};
	   		   \node at (2.23,2.3) {$c_h$};
	   		     \node at (0.3,2.2) {$\color{blue!80!black}{R}$};
	\node at (-2,6) {$p>0$};
	\end{scope}
	\begin{scope}[xshift = 3.75cm]
	\draw[line width = .2mm] (-2.5,2) -- (2.44,2) node[right]{\small $t=1$} ;
	\draw[line width = .5mm,blue!80!black] (-1.53,2) -- (1.53,2);
	\draw[->,line width = .3mm] (0,0.03) -- (0,6)node[right]{ $t$};
	\draw [line width = .3mm,blue!30!white, fill,fill opacity = 0.1] (0,0)..controls(-.2,1.3) and (-1.2,1.7).. (-1.5,2) ..controls(-1.3,2.2) and (-.7,2.8).. (0,3)..controls(.7,2.8) and (1.3,2.2)..(1.5,2)..controls(1.2,1.7) and (.2,1.3).. (0,0);
		  \draw[line width = .5mm,snake = zigzag,red!90!black] (-2.5,0) -- (2.44,0) node[right] {$\color{black}{x}$};
		  \node[rotate = 90] at (-1.53,2) {$-$};
		  \node[rotate = 90] at (1.53,2) {$-$};
		    \node at (-1.7,2.35) {$-c_h$};
		  \node at (1.53,2.3) {$c_h$};
	\node at (-2,6) {$p<0$};
	 \node at (0.3,2.2) {$\color{blue!80!black}{R}$};
	\end{scope}
	
	

	\end{tikzpicture}
	\caption{The lower tip of $D_R$ for the region $R= [-c_h, c_h] \times {\bf R}^{d-2}$ contacts the singularity (where we omit the orthogonal directions due to translational invariance). For a super-horizon region $R$, i.e. one with $c>c_h$, its causal development $D_R$ ceases to be complete. Amongst the $d-1$ possible choices for the compact direction of $R$, at least one behaves as in the contracting case of the right figure.}
	\label{causaldom}
\end{figure}

The basic building blocks for a  reconstruction formula such as  \eqref{hkll} are the causal domains in the CFT spacetime, whose existence is not guaranteed in spacetimes with singularities. To see this, we shall  profit from the symmetries of the problem and  consider spacelike  regions of the form $R= [-c, c] \times {\bf R}^{d-2}$. With no loss of generality, we can place these regions  at fixed time $t=1$, since we may use a scaling argument to recover the behavior of  other time slices. We shall denote by $p$ the Kasner exponent along the compact direction of R. The associated causal domains, $D_R$, are the product of a $(1+1)$-dimensional causal domain oriented in the $x$ direction and ${\bf R}^{d-2}$. When drawn in the Kasner coordinates, the $(1+1)$-dimensional factors appear curved inwards or outwards depending on the sign of $p$. At any rate, the important fact about these causal domains is their intersection with the singularity for a sufficiently large value of $c$. This happens for $c> c_h =(1-p)^{-1}$, as we indicate in Figure \ref{causaldom} for both possible signs of $p$. 

A larger region $R$ supports a formally  larger operator subalgebra, and therefore boasts a potentially  larger reconstruction power.  Evidently, the presence of the singularity puts a sharp limit to any reconstruction method tailored to the causal domains $D_R$. In fact, as we show in the next section, there is an ultimate limit to Green-function reconstruction, even if we summon the complete operator algebra on a non-compact Cauchy slice.

\section{Maximal causal reconstruction}
\noindent

\begin{figure}[h]
	\centering
	\begin{tikzpicture}[scale = 1]

	\draw [line width = .5mm, mint] plot [smooth, tension=1] coordinates { (0,0) (-3.5,3.5)};
	
	\draw [line width = .3mm, gray,dashed] plot [smooth, tension=1] (0,7)..controls (-.75,5.5) and (-1.4,1)..(-1.5,0);
	\draw [line width = .3mm, gray,dashed] plot [smooth, tension=1] (0,7)..controls (-1.5,5)and (-2.9,1)..(-3,0);
	\draw [line width = .3mm, gray,dashed] plot [smooth, tension=1] (0,7)..controls (-2.25,4.5)and (-4.4,1)..(-4.5,0);
	\draw [line width = .3mm, gray,dashed] plot [smooth, tension=1] (0,7)..controls (-3,4) and (-5.9,1)..(-6,0);
	
	\draw [line width = .3mm, gray,dashed] plot [smooth, tension=1] (-7,0)..controls (-5.5,.75) and (-1,1.4)..(0,1.5);
	\draw [line width = .3mm, gray,dashed] plot [smooth, tension=1] (-7,0)..controls (-5,1.5)and (-1,2.9)..(0,3);
	\draw [line width = .3mm, gray,dashed] plot [smooth, tension=1] (-7,0)..controls (-4.5,2.25)and (-1,4.4)..(0,4.5);
	\draw [line width = .3mm, gray,dashed] plot [smooth, tension=1] (-7,0)..controls (-4,3) and (-1,5.9)..(0,6);
	
	\draw[line width = .7mm,red!90!black] (-7,0)   -- (0,7);
	\draw[line width = .8mm, snake = zigzag,red!90!black]    (0,0) -- (-7,0); 
	
	\draw[line width = .5mm] (0,7) -- (0,0);
	
	
	\node[rotate = 45] at (-3.7,3.7) {$t=+\infty$};
	\node at (-3.5,-.4) {$t=0$};
	\node[rotate = 90] at (.3,3.2) {$z=0$};
	
	\node at (-1.3,3.5) {\Huge {\RomanNumeralCaps{1}}};
	\node at (-3.3,1.45) {\Huge {\RomanNumeralCaps{2}}};

	\end{tikzpicture}
	\caption{The causally reconstructible region \RomanNumeralCaps{1} is separated from region \RomanNumeralCaps{2} by the scale invariant $t=z$ hypersurface.}
	\label{Penrosediag2}
\end{figure}
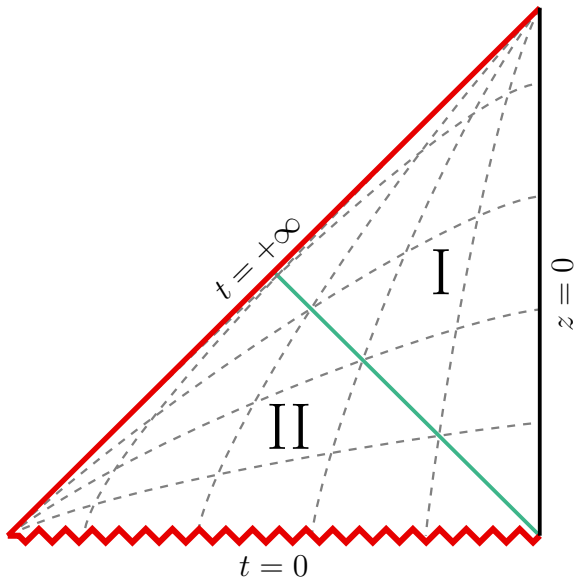

Even if we may consider regions $R$ with $c>c_h$, so that the corresponding causal developments $D_R$ are not defined, we may
ask what information can be obtained from the Heisenberg flow of operators sitting at $t>0$. In other words, what reconstruction power is held by the union of all critical domains of the form $[-c , c] \times {\bf R}^{d-2}$ for arbitrary real values of $c$. Even more optimistically, we may ask for the causal reconstruction power of the full operator algebra at $t>0$. The answer to this question is simple: causal reconstruction extends to the full $t>z$ region \RomanNumeralCaps{1} in Figure \ref{Penrosediag2}, i.e. the northeast of the Penrose diagram. 

In order to show this, we simply argue for the existence of the appropriate Green function on region \RomanNumeralCaps{1}, with initial data on the holographic boundary. For simplicity, we consider a scalar generalized free field $\mathcal{O}(t,\vec{x})$ in Kasner, and its dual bulk field $\phi(t,z,\vec{x})$ at leading large-$N$ 
\begin{equation}
(\Box - m^2 )\; \phi(t,z,\vec{x}) \,=\, 0\;.
\end{equation}

The ${\bf{R}}^{d-1}$ translational isometries guarantee that we can diagonalize the wave operator in the Fourier basis
\begin{equation}
\phi_{\vec{k}}(t,z) \,=\, \int d^{d-1} \vec{x} \;e^{-i\vec{k}\cdot\vec{x}} \;\phi(t,z,\vec{x})\;,
\end{equation}
i.e. the dynamics of each of these non-local modes is effectively two-dimensional
\begin{equation}\label{twodynamics} 
\left(\partial^2_z + \dfrac{1-d}{z}\,\partial_z -  \dfrac{1}{t}\partial_t\left(t\,\partial_t\right)- \sum_{i=1}^{d-1} t^{-2p_i}\,(k_i)^2- \dfrac{m^2}{z^2} \right)\; \phi_{\vec{k}}(t,z)  \,=\, 0\;.
\end{equation}

We can redefine the field mode $\,{\Phi}_{\vec{k}}(t,z) = z^{\frac{1-d}{2}}\,t^{\frac{1}{2}}\, \phi_{\vec{k}}(t,z)\,$ so that \eqref{twodynamics} becomes more revealing\footnote{The solution to \eqref{twodynamics} as a sum over momentum modes in the $z$-coordinate is not advantageous for causal reconstruction purposes, since these modes probe the bulk too deep (region \RomanNumeralCaps{2}) in order for the problem to have a solution in terms of data on the holographic boundary.}
\begin{equation}\label{twodynamics2}
\left(\Box^{(0)} - \, V_{\vec{k}}\,(t,z) \right)\; \Phi_{\vec{k}}(t,z)  \,=\, 0 \;, 
\end{equation}
where $\,\Box^{(0)} = -\partial_t^2 + \partial_z^2\,$ is the two-dimensional wave operator in flat spacetime, and the effective potential corresponds to
\begin{equation}
V_{\vec{k}}\,(t,z) \,=\, \dfrac{m_c^2}{z^2} \,+\, \dfrac{1}{4t^2} \,+\, \sum_{i=1}^{d-1} \dfrac{k_i^2}{t^{2p_i}}\;,
\end{equation}
once the renormalized mass $\,m_c^2 = m^2 + (d^2-1)/4\,$ is defined. The advantage of effectively reducing the number of dimensions to two is that we can switch space and time just by adding an overall minus sign, and this way reformulate the problem as an initial-value problem in an open set of flat spacetime. Assuming analyticity in $m_c$ and $\vec{k}$ for the corresponding retarded Green function, it is possible to formally build it from a series expansion with the bare free massless retarded Green function. It is not hard to see that each term in this expansion will actually have the appropriate support.

As a consequence, the desired ``spacelike retarded'' Green function can be assumed to exist in two dimenisons, solving
\begin{equation}
\mathcal{D}'_{\vec{k}}\; G_{\vec{k}}^{(-)}(t,z;t',z')  \,=\,  \delta(z-z')\,\delta(t-t')\;, \label{Greenretarded}
\end{equation}
where $\mathcal{D}'_{\vec{k}} \, = \, \Box^{(0)}  \,-\, V_{\vec{k}} $ is the differential operator in \eqref{twodynamics2} acting over the prime coordinates, with $G_{\vec{k}}^{(-)}$ supported only when $z-z' > 0$ together with $(z-z')^2-(t-t')^2 \geq 0$.

	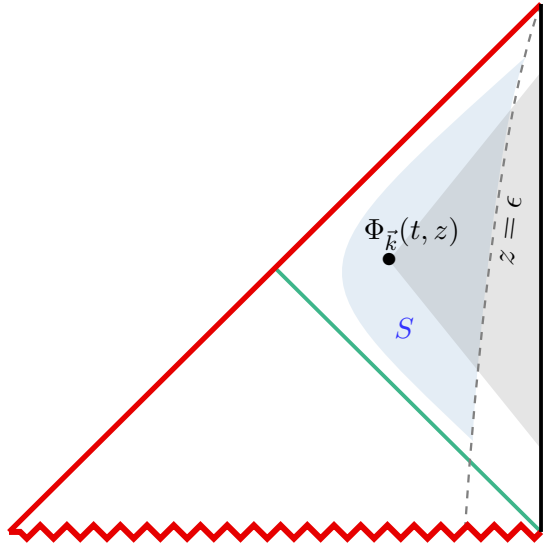
\begin{figure}[h]
		\centering
	\begin{tikzpicture}[scale = 1]

		\path [line width = .3mm,color=paleaqua,fill,fill opacity = 0.4] (-.89,1.2)..controls (-.95,1)and (-.6,5.5)..(-0.2,6.3)..controls (-3.3,3.5).. (-.89,1.2);
		
		\draw [line width = .5mm, mint] plot [smooth, tension=1] coordinates { (0,0) (-3.5,3.5)};
	
		\path [line width = .3mm,color=gray,fill,fill opacity = 0.2] (-2,3.6)-- (0,6.1)--(0,1.1)--(-2,3.6);

	\draw [line width = .3mm, gray,dashed] plot [smooth, tension=1] (0,7)..controls (-.6,5.5) and (-.95,1)..(-1,0);
	
	\draw[line width = .7mm,red!90!black] (-7,0)   -- (0,7);
	\draw[line width = .8mm, snake = zigzag,red!90!black]    (0,0) -- (-7,0); 
	
	\draw[line width = .5mm] (0,7) -- (0,0);
	

	\node at (-2,3.6) {$\bullet$};
	\node[rotate = 83] at (-.4,4) {$z = \epsilon$};
    \node at (-1.7,3.95) {\small{$\Phi_{\vec{k}}(t,z)$}};
    
    \node at (-1.8,2.7) {\small{$\color{blue!80!white}{S}$}};
	
	\end{tikzpicture} 	\caption{The support of the Green function $G_{\vec{k}}^{(-)}$ in gray for a given point in region \RomanNumeralCaps{1} and its intersection with the $\epsilon$-regularized boundary. The blue volume $S$ is conveniently taken to apply Gauss' law to the point in question.}\label{domain}\end{figure}

Having argued for the existence of such a Green function, it is straightforward at this point to write down an expression for the general solution in region  \RomanNumeralCaps{1} in terms of boundary data. We consider a regularized boundary at $\epsilon$ coordinate distance and integrate over the blue region $S$ in Figure \ref{domain} to get
\begin{equation}
\Phi_{\vec{k}}(t,z) \big|_{\text{\RomanNumeralCaps{1}}} \,= \,\int_S dt'\,dz'\; \delta(z-z')\,\delta(t-t')\, \Phi_{\vec{k}}(t',z') \; = \int_S dt'\,dz'\; \Phi_{\vec{k}}(t',z')\; \mathcal{D}'_{\vec{k}}\; G_{\vec{k}}^{(-)}(t,z;t',z') \;,
\end{equation}
and now integrate by parts, using Gauss' law, to get 
\begin{equation}
\Phi_{\vec{k}}(t,z) \big|_{\text{\RomanNumeralCaps{1}}} \,= \,\int_0^\infty dt'\, \left(G_{\vec{k}}^{(-)}(t,z;t',\epsilon)\; \partial_{z'} \Phi_{\vec{k}}(t',\epsilon) \,-\, \Phi_{\vec{k}}(t',\epsilon)\; \partial_{z'} G_{\vec{k}}^{(-)}(t,z;t',\epsilon)\right)\;. \label{fielddef}
\end{equation}
where we used that the remaining volume integral vanishes since $ \mathcal{D}'_{\vec{k}}\,\Phi_{\vec{k}} = 0$, and moreover that both $\partial S \cap \text{supp}(G^{(-)}_{\vec{k}})$ and $\partial S \cap \text{supp}(\partial G^{(-)}_{\vec{k}})$ lie on the $\epsilon$-regularized boundary.

We can move the boundary to spatial infinity by taking $\epsilon\rightarrow0$. The extrapolate dictionary dictates that this mode will asymptote to the corresponding mode of the dual CFT field
\begin{equation}
\Phi_{\vec{k}}(t,\epsilon) \,\sim \,\epsilon^{\frac{1}{2}+ \nu}\; \mathcal{O}_{\vec{k}}(t)\;,
\end{equation} 
where $\nu^2 = d^2/4 + m^2$ and the unusual scaling is simply a consequence of the previous field redefinition. Likewise, since the Green function satisfies the homogeneous version of \eqref{Greenretarded} for any point at finite $z$ as we take $\epsilon\rightarrow 0$, it will have both a normalizable and a non-normalizable component
\begin{equation}
G_{\vec{k}}^{(-)}(t,z;t',\epsilon)\, \sim \, (2\nu)^{-1} \,\left(\epsilon^{\frac{1}{2}+ \nu} \;L_{\vec{k}}(t,z;t') \,+\, \epsilon^{\frac{1}{2}- \nu} \;K_{\vec{k}}(t,z;t')\right) \;.
\end{equation} 

The normalizable part $L_{\vec{k}}$ gives terms in \eqref{fielddef} proportional to $\epsilon^{2\nu}$, in contrast with the $\epsilon$-independent terms that come from $K_{\vec{k}}$. Accordingly,
\begin{equation}
\Phi_{\vec{k}}(t,z) \big|_{\text{\RomanNumeralCaps{1}}}  \,=\, \int_0^\infty dt' \,K_{\vec{k}}(t,z;t')\; \mathcal{O}_{\vec{k}}(t')\;,\label{fielddef2}
\end{equation}
which, after undoing the field redefinition and the Fourier transform, results in 
\begin{equation}
\phi(t,z,\vec{x})\big|_{\text{\RomanNumeralCaps{1}}} \,=\, t^{-\frac{1}{2}} \;z^{\frac{d-1}{2}}\;\int \dfrac{d^{d-1}\vec{k}}{(2\pi)^{d-1}} \,e^{i\vec{k}\cdot\vec{x}} \;\int_0^\infty dt' \,K_{\vec{k}}(t,z;t')\; \mathcal{O}_{\vec{k}}(t')\;.\label{fielddef5}
\end{equation}

The $t'$ integral has compact integration range, hence we expect that it converges for all $\vec{k}$, and moreover that it decays fast enough for large $|\vec{k}|$, so that we can swap both integrals to re-express \eqref{fielddef2} in the more familiar form 
\begin{equation}
\phi(t,z,\vec{x})\big|_{\text{\RomanNumeralCaps{1}}}\, =\,\int_{t'>0} dt'\,\int d^{d-1}\vec{x}'\, K(t,z,\vec{x},t',\vec{x}') \;\mathcal{O}(t',\vec{x}') \;,\label{fielddef6}
\end{equation}
where the smearing function is, up to some multiplicative factors, the inverse Fourier transform of $K_{\vec{k}}$, understood in the distributional sense
\begin{equation}
K(t,z,\vec{x};t',\vec{x}') \,=\, t^{-\frac{1}{2}} \;z^{\frac{d-1}{2}}\;\int \dfrac{d^{d-1}\vec{k}}{(2\pi)^{d-1}}\, e^{i\vec{k}\cdot(\vec{x}-\vec{x}')}\;K_{\vec{k}}(t,z;t')\;.
\end{equation}
These considerations extend trivially for $t<0$, the causal reconstruction horizon being $z=-t$ in this case. An analogous construction exists in terms of a faithful coordinate system covering region I alone (see Appendix 1). At any rate, we see that causal reconstruction is completely explicit outside the horizons $z=|t|$, either to the past or the  future of the singularity. 

\section{Modular Supremacy}
\noindent 

In this section we turn to the study of entanglement wedges in the Kasner-AdS state. These conform the geometrical data for modular reconstruction, and our main question is wether the reconstruction horizon $z=|t|$ is modular-traversable or not. In other words, we are interested in detecting HRT surfaces which penetrate beyond $z=|t|$ in the bulk. 

Our choice of boundary region $R = [-c, c]\,\times\, {\bf R}^{d-2}$ is essentially motivated by symmetries, and its convenience for calculating extremal surfaces resides in the fact that its HRT will factorize as $\,\chi_R\, = \,\gamma \; \times \; \mathbf{R}^{d-2}\,$, where $\gamma$ is a spacelike trajectory on the $(t,z,x)$-plane, and the ${\bf{R}}^{d-2}$ factor corresponds to the orthogonal directions. 

We can therefore restrict to the family of surfaces that factorize in the same manner, so that the volume functional reduces to
\begin{gather}
\text{Vol}_{d-1}\, = \,\int_{s_0}^{s_f} ds\, \int d^{d-2}x_j  \; \sqrt{h} \;=\, \int d^{d-2}x_j \, \int_{s_0}^{s_f} \,ds\, \dfrac{t^{1-p}}{z^{d-1}} \;\sqrt{(z')^2-(t')^2+ t^{2p}(x')^2} \;,
\end{gather}
where $s$ is a parameter along the trajectory, and prime is $d/ds$. The $x_j$ integrals yield the contribution from the orthogonal directions, which needs to be regularized
\begin{equation}\label{geodesic}
\text{Vol}_{d-1} \,=\,  \text{Vol}_{d-2}\; \int_{\tau_0}^{\tau_f} d\tau\;.
\end{equation}

Since the HRT extremizes \eqref{geodesic}, its first factor $\gamma$ will correspond to a spacelike geodesic in the effective background
\begin{equation}
d\tau^2 \,=\, \dfrac{t^{2(1-p)}\;dz^2\,-\,t^{2(1-p)}\;dt^2 \,+\, t^2 \;dx^2}{z^{2(d-1)}}\;,
\end{equation}
subject to the boundary conditions
\begin{gather}
z\,(\tau_0) \,=\, z\,(\tau_f) \,=\, \epsilon \label{one}\\
t\,(\tau_0) \,=\, t\,(\tau_f) \,=\, 1 \label{two}\\
x\,(\tau_0) \,=\, -c  \hspace{.5cm},\hspace{.5cm} x\,(\tau_f) \,=\, c \;,\label{three}
\end{gather}
where $\epsilon$ is the bulk regulator that brings the boundary to finite spatial distance. 

The evaluation of the on-shell volume for the HRT is just  
\begin{equation}
\text{Vol}_{d-1}\,[\chi_R] \,=\,  \text{Vol}_{d-2}\; 2\tau_{\text{max}}\;.
\end{equation}
where $2\tau_{\text{max}} = 2\tau_{\text{max}} (\epsilon,\lambda)$ is the regularized proper length of the geodesic $\gamma$.

In affine parametrization, the equations of motion read
\begin{gather}
{z}''  \,=\, \dfrac{d-1}{z}\;\left((z')^2 +(t')^2-t^{2p}\;(x')^2 \right) \,-\, \dfrac{2(1-p)}{t}\; t'\;z' \label{zeta}\\
{t}''  \,=\, \dfrac{2(d-1)}{z} \;t'\;z' \,-\, \dfrac{(1-p)}{t} \;\left((t')^2+(z')^2\right) \,-\, t^{2p-1}\;(x')^2 \label{time}\\
x'' \,=\, -\dfrac{2}{t}\;x'\;t' \,+\,  \dfrac{2(d-1)}{z}\; x'\; z' \;.\label{ekis}
\end{gather}

Our first observation is that there must be a turning point for which $t'\,(\tau_\star) = 0$, as required by the boundary condition \eqref{two}. The second is that every point with $ t'\,(\tau_\star)= 0$ is a maximum of $t(\tau)$ and there is thus only one such turning point. This implies that $t(\tau) \geq 1 $ throughout the geodesic. By $x\rightarrow -x$ inversion symmetry, the turning point must be at $x\,(\tau_\star)= 0$, and thus it must also have $z'(\tau_\star)=0$, being a maximum of  $z\,(\tau)$ too\footnote{This is not true if there was a twofold degeneracy on the length, which we assume not to be the case here.}. We will conveniently choose the affine parameter to set the turning point at $\tau_\star = 0$.

Given these considerations, we can reformulate the geodesic problem in terms of the turning conditions
\begin{gather}
z\,(\tau = 0) = z_\star \hspace{1cm} z'\,(\tau = 0) = 0\label{one1}\\
t\,(\tau = 0) = t_\star \hspace{1.2cm} t'\,(\tau = 0) = 0\label{two1}\\
\hspace{.3cm}x\,(\tau = 0) = 0  \hspace{1.2cm} x'(\tau = 0) = x'_\star \;.\label{three1}
\end{gather}

The affine parametrization fixes $x'_\star = z_\star^{d-1}\;t_\star^{-1}\,,$ which means that we can use the two coordinates of the tip, $z_\star$ and $t_\star$, to specify the geodesic. Generically, such geodesics will not end at $t=1$ in the boundary, contrary to what we requested previously in \eqref{two}. We can nevertheless solve for generic $z_\star$ and $t_\star$, and obtain the time  $t_b$ at which the endpoints of the geodesic touch the regularized boundary. Then, we can rescale everything by $t_b$ so that the new geodesic intersects the boundary at $t=1$, as we demanded for our choice of region $R$. 

In other words, the only significant parameter that survives this rescaling and labels unequivocally the geodesics that end at $t=1$ is the ratio $\lambda \,=\, t_\star\,/\,z_\star\,$. This is in accordance with the fact that the only boundary parameter we can play with is the length $c$ in the $x$-direction. It is reasonable to expect a one-to-one correspondence between both of them, of the form $c \,=\, c\,(\lambda)$.

\cleardoublepage

\subsection{Qualitative behavior of the entanglement wedges}
\noindent

We are now in the position to examine the existence of HRTs in Kasner-AdS that explore region \RomanNumeralCaps{2}, and see the shape of the corresponding entanglement wedges. Since our choice of boundary regions $R$ allows for a parametrization of the corresponding HRTs in terms of $\lambda \,=\,t_\star/z_\star\,$, the interesting HRTs are the ones with $\lambda \,< \,1\;$. In particular, we will next show that these do not correspond to regions $R$ which are super-horizon necessarily. 

Unfortunately, we lack of analytical solutions to the geodesic equations, so all the following results will have a numerical origin. Remarkably, the qualitative features of the HRTs seem to be almost insesitive to the exact number of boundary spacetime dimensions\footnote{Kasner solutions to the system $\,\sum_j \,p_j = \sum_j \,p_j^2 =1\,$ only exist when the the total number of spacetime dimensions $d$ is greater than $3$. The value of the smallest Kasner coefficient $p_{\text{min}}$ is bounded by  bellow by $-1+2/(d-1)$. We will not contemplate Milne-type solutions in this article, that is. some Kasner coefficient being one (and all the others vanishing), since these correspond to a boundary flat metric.} $d$ as well as to the value of the Kasner coefficient $p$ in the compact direction of $R$. For this reason, we can pick $p \,=\, -1/2\,$ in $d = 5$ as a good representative of what happens in general. This choice satisfies the Kasner constraints with all the orthogonal directions being $p_j = 1/2$ ($j =1,2,3$).

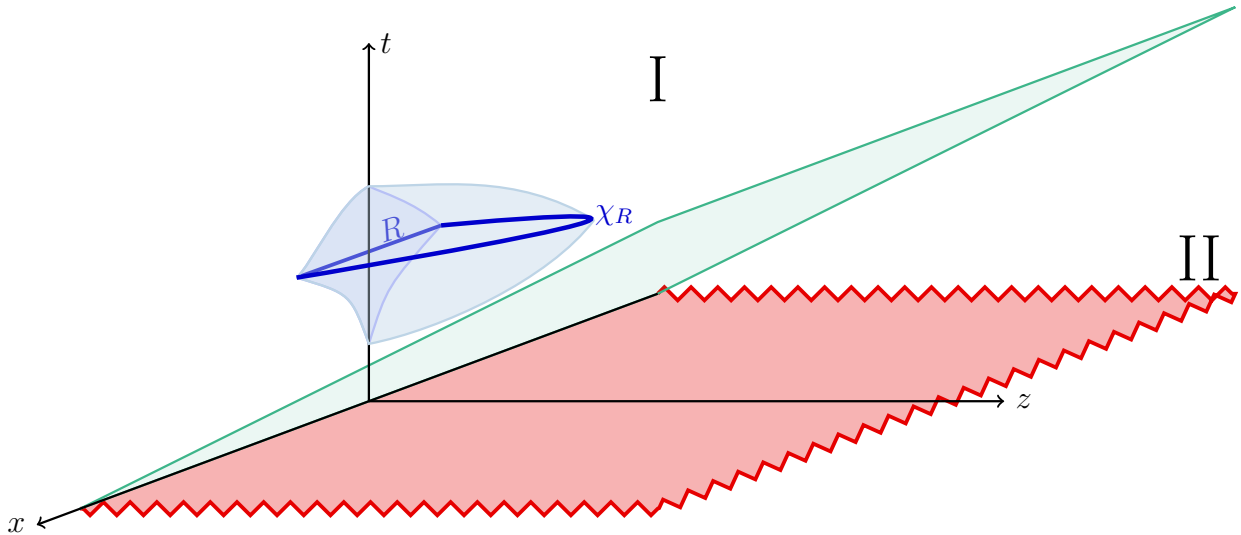
\begin{figure}[h]
	\centering
	\begin{tikzpicture}[scale = .95]
	
	\draw[line width = .5mm,red!90!black, fill,fill opacity = 0.3,snake = zigzag] (0,-3.5)--(8,-3.5)--(0,-6.5)--(-8,-6.5);
	\draw[line width = .3mm,mint, fill,fill opacity = 0.1] (-8,-6.5)--(0,-3.5)--(8,.5)--(0,-2.5)--(-8,-6.5);
	
		\draw[line width = .3mm,blue!30!white, fill,fill opacity = 0.2] (-5,-3.275)..controls (-4.8,-3.1)and(-4.4,-2.2)..(-4,-2)..controls(-3.8,-2.1)and(-3.4,-2.2)..(-3,-2.547)..controls(-3.75,-3.4)..(-4,-4.2)..controls(-4.25,-3.5)..(-5,-3.275);
	
	\draw[line width = .5mm,blue!80!black] (-5,-3.275)--(-3,-2.547);
	
	\draw[->,line width = .3mm]    (-4,-5) -- (4.8,-5)node[right]{$z$}; 
	
	\draw[->,line width = .3mm] (-4,-5) -- (-4,0)node[right]{$t$};
	
	\draw[->,line width = .3mm] (0,-3.5)--(-8.6,-6.725) node[left]{$x$};

	\node[rotate = 18] at (-3.7,-2.6) {$\color{blue!80!black}{R}$};
	\draw[line width = .3mm,paleaqua, fill,fill opacity = .4] (-5,-3.275)..controls (-4.8,-3.1)and(-4.4,-2.2)..(-4,-2)..controls(-3,-2) and (-2,-1.8)..(-.88,-2.47);
	\draw[line width = .3mm,paleaqua, fill,fill opacity = 0.4] (-5,-3.275)..controls (-4.25,-3.5)..(-4,-4.2)..controls(-3,-4) and (-1.5,-3.3)..(-.88,-2.47);
	
	\draw [line width = .6mm,blue!80!black] plot [smooth, tension=1] coordinates { (-5,-3.275)(-1,-2.5)(-3,-2.547)};

	\node at (0,-0.5) {\Huge {\RomanNumeralCaps{1}}};
	\node at (7.5,-3) {\Huge {\RomanNumeralCaps{2}}};
	
	\node at (-.6,-2.4) {$\color{blue!80!black}{\chi_R}$};

	
	\end{tikzpicture}
	\caption{The entanglement wedge $EW_R$ for a sub-critical $c\,<\,c_{\text{cr}}$ boundary region $R$ in the contracting branch ($p<0$) stays in region \RomanNumeralCaps{1}. Its boundary causal domain $D_R$ is complete since the region $R$ is sub-horizon length. The entanglement wedge has a spacelike profile due to caustics.}
	\label{figone}
\end{figure}

It is clearly more illuminating to think in terms of the length $c$ of the boundary region $R$ rather than in terms of $\lambda$, so we will explain our results this way. We begin with the zero-length region and gradually start increasing its length $c$. For small enough regions $R$, which we shall refer to as sub-critical, the situation is shown in Figure \ref{figone}, where the HRT $\chi_R$ remains in region \RomanNumeralCaps{1}. Besides, in this regime, $R$ has sub-horizon length $c<c_h$.

For sub-critical regions $R$, the entanglement wedge $EW_R$ is contained within region \RomanNumeralCaps{1}. The reason is that $\, EW_R \,\subset \,U \,\times\, \mathbf{R}^{d-1}\, \subset \, \text{region \RomanNumeralCaps{1}}\,$, where $U$ is the region in the $(t,z)$-plane delimited by the light rays that emanate from the tip of the geodesic $(t_\star,z_\star)$ and reach the boundary, i.e. $U$ is exactly the support of the spacelike retarded Green function of Figure \ref{domain}.

\begin{figure}[h]
	\centering
	\includegraphics[scale = .7]{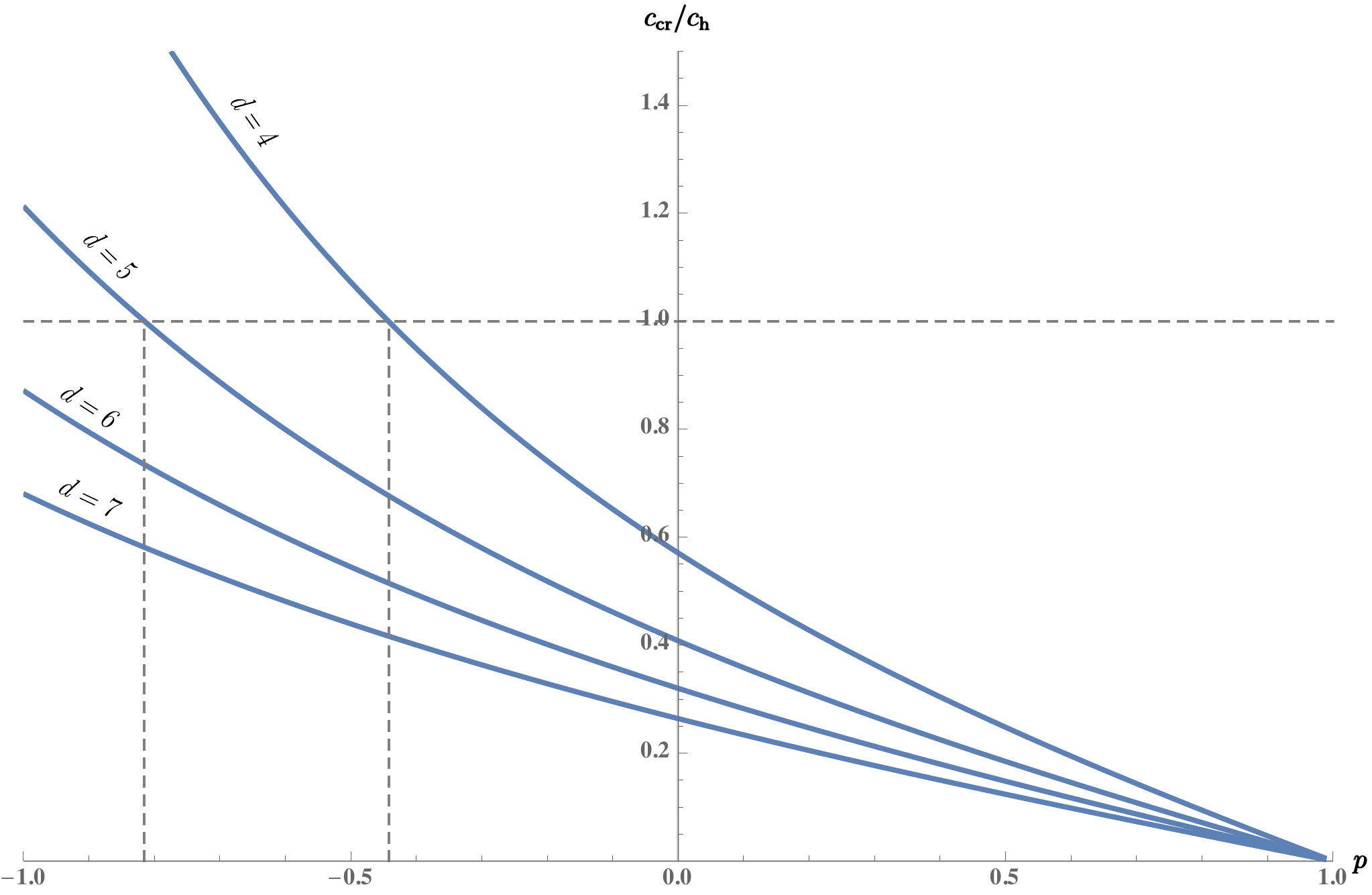}
	
	\caption{The critical length $c_{\text{cr}}$ is smaller than the horizon length $c_h$  for any Kasner-AdS, for which $p_{\text{min}}\,(d)\, =\, -1\,+\,2/(d-1)\,.$ }
	\label{figuracritica}
\end{figure}

Eventually, the length of $R$ reaches the critical value $\,c= c_{\text{cr}}\,$ for which the tip of the HRT $\chi_R$ contacts the hyperplane that separates region \RomanNumeralCaps{1} from \RomanNumeralCaps{2}. The same argument as above assures that, still at this point, $EW_R$ does not intersect region \RomanNumeralCaps{2}. For any Kasner solution, the critical length in any of the $(d-1)$-directions $c_{\text{cr}}\,(d,p)$ is smaller than the cosmological horizon length $c_h\,$, as we show in Figure \ref{figuracritica}. In particular, for $p=-1/2\,$ in $d=5$, it is around $72 \%$ of $c_h$.

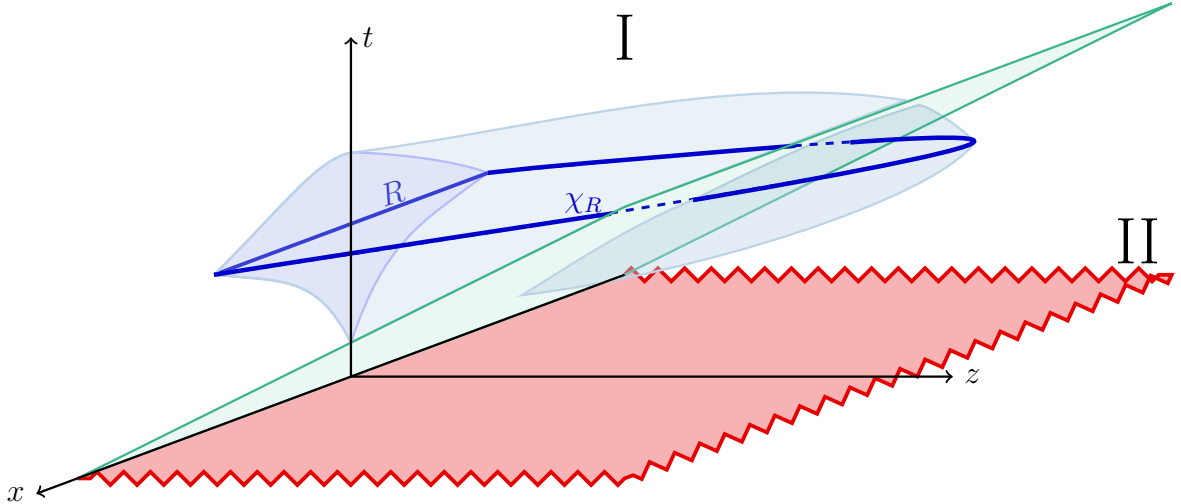
\begin{figure}[h]
	\centering
	\begin{tikzpicture}[scale = .9]

	\draw[line width = .5mm,red!90!black, fill,fill opacity = 0.3,snake = zigzag] (0,-3.5)--(8,-3.5)--(0,-6.5)--(-8,-6.5);

		\draw[line width = .3mm,blue!30!white, fill,fill opacity = 0.2] (-6,-3.5)..controls(-5,-2.5)and(-4.5,-1.8)..(-4,-1.7)..controls(-3.5,-1.7)and (-2.5,-1.8)..(-2,-2)..controls(-3,-2.75)and(-3.5,-3)..(-4,-4.5)..controls(-4.5,-3.5) and(-5,-3.6)..(-6,-3.5);
	\draw[line width = .5mm,blue!80!black] (-6,-3.5)--(-2,-2);

	\node[rotate = 18] at (-3.4,-2.3) {$\color{blue!80!black}{R}$};


	\draw[line width = .3mm,paleaqua, fill,fill opacity = 0.3] (-6,-3.5)..controls(-5,-2.5)and(-4.5,-1.8)..(-4,-1.7)..controls(-1.5,-1.4)and (1.5,-.5)..(4.1,-.93)--(0,-2.5)--(-2.5,-3.75)--(-4,-4.5)..controls(-4.5,-3.5) and(-5,-3.6)..(-6,-3.5);
	
	\draw[line width = .3mm,paleaqua, fill,fill opacity = 0.4] (5.12,-1.55)..controls(4.85,-1.25)and (4.4,-1) ..(4.3,-1)..controls(3,-1.4)and (2,-1.9) .. (1,-2.4)..controls(2.5,-2.15) and(4.5,-1.8).. (5.12,-1.55);

	\draw [line width = .4mm,blue!80!black,dashed] plot [smooth, tension=1] coordinates { (-6,-3.5)(5,-1.6)(-2,-2)};
	\draw [line width = .6mm,blue!80!black] plot [smooth, tension=1] coordinates { (-6,-3.5)(-2.5,-2.95)(-.2,-2.6)};
	\draw [line width = .6mm,blue!80!black] plot [smooth, tension=1] coordinates { (2.5,-1.6)(-.5,-1.85)(-2,-2)};

		\draw[line width = .3mm,mint, fill,fill opacity = 0.1] (-8,-6.5)--(0,-3.5)--(8,.5)--(0,-2.5)--(-8,-6.5);
		
		\draw[line width = .3mm,paleaqua, fill,fill opacity = 0.4] (5.12,-1.55)..controls(4.5,-2.4) and(1,-3.5)..(-1.5,-3.8)..controls(-1,-3.5) and(0,-2.8) .. (1,-2.4)..controls(2.5,-2.15) and(4.5,-1.8).. (5.12,-1.55);
	
	\draw [line width = .6mm,blue!80!black] plot [smooth, tension=1] coordinates { (3.3,-1.55)(5,-1.6)(1,-2.4)};
	\node at (0,0) {\Huge {\RomanNumeralCaps{1}}};
	\node at (7.5,-3) {\Huge {\RomanNumeralCaps{2}}};
	
	\node at (-.6,-2.45) {$\color{blue!80!black}{\chi_R}$};
	
	\draw[->,line width = .3mm]    (-4,-5) -- (4.8,-5)node[right]{$z$}; 
	
	\draw[->,line width = .3mm] (-4,-5) -- (-4,0)node[right]{$t$};
	
	\draw[->,line width = .3mm] (0,-3.5)--(-8.6,-6.725) node[left]{$x$};

	
	\end{tikzpicture}
	\caption{The entanglement wedge $EW_R$ for a super-critical $c\,>\,c_{\text{cr}}$ but still sub-horizon boundary region $R$ in the contracting branch ($p<0$) probes region \RomanNumeralCaps{2}. Its boundary causal domain $D_R$ is complete, and so is the entanglement wedge $EW_R$.}
	\label{figtwo}
\end{figure}

Slightly above the critical length, the region $R$ becomes super-critical while preserving its sub-horizon character. As we show in Figure \ref{figtwo}, the HRT $\chi_R$ will finally enter region \RomanNumeralCaps{2}, and the entanglement wedge $EW_R$ will still be complete. This last property is a consequence of the fact that $\, EW_R \,\subset \,[0,z_\star]\,\times\, D_z\,\times\, \mathbf{R}^{d-2}\,$, where $D_z$ is the $(t,x)$-diamond (as the ones in Figure \ref{causaldom}) for the interval with $x(z)$-length and placed at constant $t(z)$ time, both of which are given by the corresponding coordinates of the HRT $\chi_R$ at that particular value of $z$. Since time $t(z)$ grows monotonically with $z$ for the HRT $\chi_R$, while the length $x(z)$ decreases monotonically, each of these causal diamonds will be complete only if the boundary diamond $D_R$ is complete, i.e. when the region $R$ has a sub-horizon size.



\begin{figure}[h]
	\centering
	\begin{tikzpicture}[scale = .86]
	
	\draw[line width = .5mm,red!90!black, fill,fill opacity = 0.3,snake = zigzag] (0,-3.5)--(8,-3.5)--(0,-6.5)--(-8,-6.5);

		\draw[line width = .3mm,blue!30!white, fill,fill opacity = 0.2] (-7,-4.25)..controls (-6,-3)and(-4.5,-1.2)..(-4,-1)..controls(-3.5,-.8)and(-2,-1.65)..(-1,-1.9)..controls(-2,-2.6)and(-2.75,-3)..(-3.25,-4.7)--(-4.5,-5.22)..controls(-4.75,-4.7) and(-6,-4.29).. (-7,-4.25);
	
	\draw[line width = .5mm,blue!80!black] (-7,-4.25)--(-1,-1.9);

	\node[rotate = 18] at (-3.1,-2.5) {$\color{blue!80!black}{R}$};
	
	\draw[line width = .3mm,paleaqua, fill,fill opacity = 0.4] (-7,-4.25)..controls (-6,-3)and(-4.5,-1.2)..(-4,-1)..controls(-1,-.5) and(4,0.3)..(6.5,-0.05)--(-0,-2.5)--(-4.8,-4.9)..controls(-5,-4.7)and(-6,-4.29).. (-7,-4.25);

\draw[line width = .3mm,paleaqua, fill,fill opacity = 0.4] (8.27,-.9)..controls(7.75,-.4) and(7,-.1)..(6.6,-0.12)..controls(4,-1.2) and(1,-2.15)..(-.3,-2.9)..controls(3.5,-2.1)and(6.5,-1.6) ..(8.27,-.9);

	\draw [line width = .4mm,blue!80!black,dashed] plot [smooth, tension=1] coordinates { (-7,-4.25)(8,-1.)(-1,-1.9)};
	
	\draw [line width = .6mm,blue!80!black] plot [smooth, tension=1] coordinates { (3.1,-1.33)(1,-1.6)(-1,-1.9)};
	
	\draw [line width = .6mm,blue!80!black] plot [smooth, tension=1] coordinates { (-7,-4.25)(-3.5,-3.55)(-1.2,-3.08)};
	
	\draw[line width = .3mm,mint, fill,fill opacity = .1] (-8,-6.5)--(0,-3.5)--(8,.5)--(0,-2.5)--(-8,-6.5);


	\draw[line width = .3mm,paleaqua, fill,fill opacity = 0.4] (8.27,-.9)..controls(7,-3) and(1,-4.5)..(-1.3,-5)..controls(-2,-5) and(-3,-5.1)..(-4.5,-5.22)..controls(-3,-4.2) and(-1.5,-3.5).. (-.3,-2.9)..controls(3.5,-2.1)and(6.5,-1.6) ..(8.27,-.9);
\draw [line width = .6mm,blue!80!black] plot [smooth, tension=1] coordinates { (-.3,-2.9)(8.06,-1)(4,-1.23)};

	\draw[line width = .3mm,blue!50!white,dashed] (-1.3,-5)..controls(-2,-5) and(-3,-5.1)..(-4.5,-5.22);
\draw[line width = .3mm,blue!50!white,dashed] (-1.3,-5)..controls(-1.75,-4.9) and(-2.5,-4.9)..(-3.25,-4.75);
	
	\draw[->,line width = .3mm]    (-4,-5) -- (4.8,-5)node[right]{$z$}; 
	
	\draw[->,line width = .3mm] (-4,-5) -- (-4,0)node[right]{$t$};
	
	\draw[->,line width = .3mm] (0,-3.5)--(-8.6,-6.725) node[left]{$x$};

	\node at (0,.5) {\Huge {\RomanNumeralCaps{1}}};
	\node at (7.5,-3) {\Huge {\RomanNumeralCaps{2}}};

	\node at (5,-1.55) {$\color{blue!80!black}{\chi_R}$};

	
	\end{tikzpicture}
	\caption{The entanglement wedge $EW_R$ for a super-critical $c\,>\,c_{\text{cr}}$ and super-horizon boundary region $R$ in the contracting branch ($p<0$) probes region \RomanNumeralCaps{2}. Its boundary causal domain $D_R$ is complete, and so is the entanglement wedge $EW_R$.}
	\label{figthree}
\end{figure}
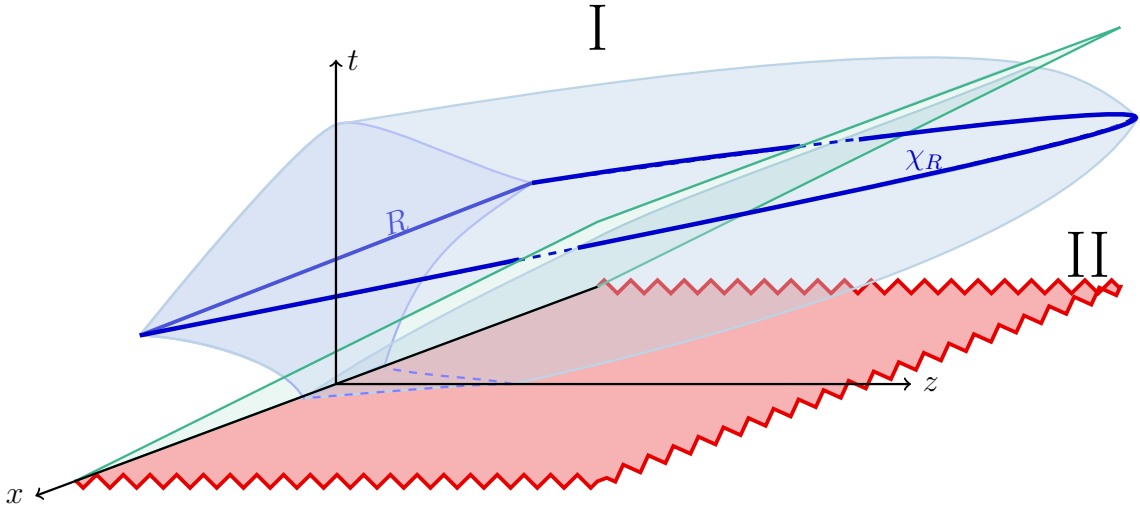

Lastly, we will reach the cosmological horizon length $c = c_h$. By the previous argument, it is at this same moment when the entanglement wedge $EW_R$ will first touch the singularity. From this point forward, the region $R$ becomes super-critical and, at the same time, has super-horizon length. The entanglement wedge $EW_R$ intersects the bulk singularity in the way shown in Figure \ref{figthree}.

\begin{figure}[h]
	\centering
	\begin{tikzpicture}[scale = 1]

	\draw [line width = .5mm, mint] plot [smooth, tension=1] coordinates { (0,0) (-3.5,3.5)};

		\draw[line width = .5mm, brown!50!white] plot [smooth, tension=1] (0,2)..controls (-2,2)and (-4.3,2.5)..(-7,0);
			\draw [line width = .3mm, gray,dashed] plot [smooth, tension=1] (-7,0)..controls (-5.5,1) and (-2.5,2)..(0,2);
	
	\draw[line width = .7mm,red!90!black] (-7,0)   -- (0,7);
	\draw[line width = .8mm, snake = zigzag,red!90!black]    (0,0) -- (-7,0); 
	
	\draw[line width = .5mm] (0,7) -- (0,0);
	

	\node at (0,2) {\color{blue!80!black}{$\bullet$}};
	\node at (0.8,2) {$\color{blue!80!black}{\lbrace R \rbrace_{c=0}^\infty}$};
	\node[rotate = 90] at (-2.03,2.03) {\small{$\bullet$}};
	\node at (-1.95,2.3) {\small $c_{\text{cr}}$};
	
	\node[rotate = 90] at (-3.05,1.95) {\small{$\bullet$}};
    \node[rotate = 15]  at (-4,1.1) {\small $\color{gray}{t=1}$};
      \node at (-4.1,2.1) {\large $\color{brown!50!white}{\Sigma}$};
	
	\node at (-3,2.2) {\small $c_h$};
	
	\node at (-1,4) {\Large {\RomanNumeralCaps{1}}};
	\node at (-3,.6) {\Large {\RomanNumeralCaps{2}}};


	\end{tikzpicture}
	
	\caption{Given the one-parameter family of regions $\lbrace R \rbrace_{c=0}^\infty$ oriented along some fixed compact direction, the corresponding one-parameter family of HRTs $\lbrace \chi_R \rbrace_{c=0}^\infty$ generates a bulk Cauchy slice $\Sigma$. Its $(t,z)$-projection is produced by the family of endpoints of the HRTs. The endpoint of the HRT for the critical region $R$ lies on the green hypersurface, while for the region $R$ of cosmological horizon size, it gets deeper into the bulk.}
	\label{hrt}
\end{figure}
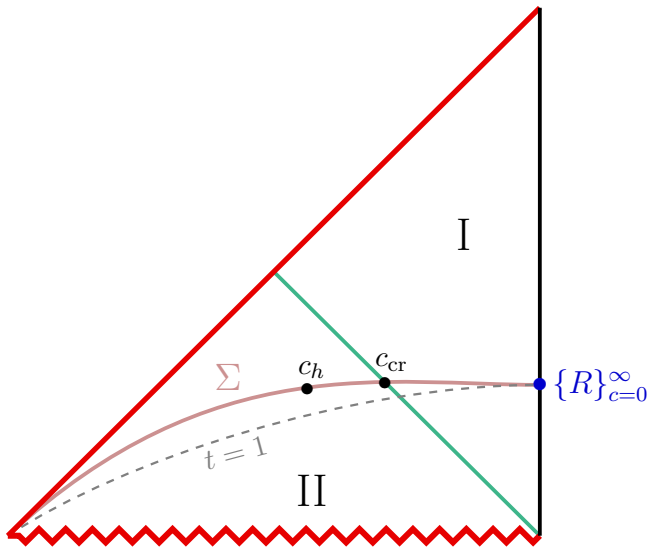

Naturally, the larger the compact direction of $R$ is, the more its HRT $\chi_R$ will explore region \RomanNumeralCaps{2}. By entanglement wedge nesting, we can generate a bulk Cauchy slice just fixing the compact direction of $R$ and considering the union of all possible HRTs, as shown in Figure \ref{hrt}.

\vspace{2cm}

\section{Conclusions}
\noindent

In this note we have mapped entanglement wedges in a holographic model, the Kasner-AdS solution, where cosmological singularities severely restrict the power of causal reconstruction methods (i.e. those based on Green's functions). The combined effect of the crunch/bang singularity at $t=0$ and the tidal singularities at $t=\pm \infty$ amounts to the existence of a horizon for causal reconstruction at $z=\pm t$, namely local bulk operators supported on the region $|t|< z$ cannot be `linearly sourced' from  single-trace CFT operators at leading order in the $1/N$ expansion. On the other hand, entanglement wedges for certain regions on a boundary slice do penetrate beyond the causal reconstruction horizon and an entire bulk Cauchy surface can be reached by nesting larger and larger regions of the CFT slice. 

We have emphasized the fact that modular reconstruction, based on a strip region R of finite transverse extent inside a fixed t-slice of the boundary, is more penetrating that the more powerful causal reconstruction using the entire CFT operator algebra on the full fixed-$t$ slice. It would be interesting to study other examples of reconstruction in the presence of cosmological singularities to assess the generality of this phenomenon. For instance, one can engineer singularities in the interior of expanding bubbles in AdS, similar to Coleman-de Luccia instabilities (cf. for example \cite{Maldacena:2010un,Harlow:2010my,Barbon:2011ta} and references therein). These states exist in the CFT defined on a sphere, and deformed by a time-dependent relevant operator. The acceleration horizon of such bubbles is a causal reconstruction horizon for the operator algebra on a Cauchy slice of the CFT sphere. On the other hand, the RT surface of a hemisphere clearly penetrates into the interior of the expanding bubble, giving a simple example of the supremacy of modular reconstruction. In this case, however, both CFT operator algebras are supported on a compact spatial region, so that the supremacy is not as neat a phenomenon as in the Kasner-AdS state.

It is interesting to  notice that the presence of the tidal singularities at $t=\pm \infty$ is crucial for the significant limitation of the causal reconstruction methods, even when supported on an entire non-compact Cauchy slice of the boundary. To illustrate this, we can replace the bare Kasner-AdS metric by a Kasner-AdS soliton metric which regularizes the $z=\infty$ limit of the spacetime. Here, we add an extra compact circle with thermal boundary conditions, introducing a `cigar' factor in the bulk metric, capped at a finite value of the radial coordinate  $z=z_0$ (cf. for example \cite{Engelhardt:2013jda}).  Causal reconstruction effectively  treats $z=z_0$ as a timelike surface with reflection boundary conditions. This is enough to solve for any bulk operator in terms of operators supported at $|t|>z$, which in turn can be mapped to the boundary by HKLL methods. This brings the  whole bulk under the control of  causal reconstruction, but only if we use the full operator algebra on a non-compact Cauchy slice of the CFT. In this sense, it is still significant that modular reconstruction will only use the operator algebra on a proper subregion of that Cauchy slice.

\section*{Acknowledgments}
\label{ackn}

We thank Javier Mart\'{\i}n Garc\'{\i}a for many discussions. This work is partially supported by the Spanish Research Agency (Agencia Estatal de Investigaci\'on) through the grants IFT Centro de Excelencia Severo Ochoa SEV-2016-0597,  FPA2015-65480-P and  
PGC2018-095976-B-C21. 
The work of M.S. supported by a predoctoral FPU grant FPU-16/00639.

\appendix

\section{Appendix: Different smearing for region I modes}
\noindent

The holographic representation of bulk approximately local degrees of freedom is highly non-unique. In this appendix, we will explicitly construct a smearing function for region \RomanNumeralCaps{1} modes which is not compactly supported on the $t$ direction and, hence, it differs from the one in section 3. To do so, we first introduce a somewhat more convenient coordinate system to describe region \RomanNumeralCaps{1}, and change the CFT frame
\begin{gather}
\begin{cases}
\;e^{\tau} \,=\, t \hspace{1.5cm} \tau \in (-\infty,\infty)\;,\\ \;e^{-\rho} \,=\, \dfrac{z}{t} \hspace{1.5cm} \rho \in (0,\infty)\;.
\end{cases}
\end{gather}
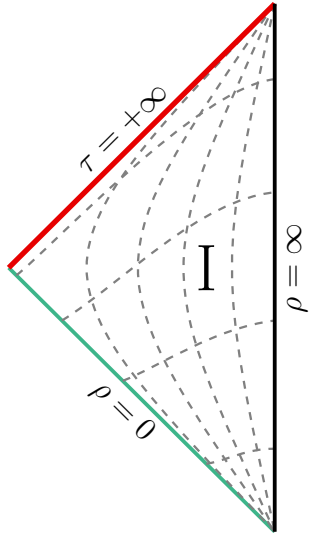
\begin{figure}[h]
	\centering
	\begin{tikzpicture}
	
	\draw [line width = .5mm, mint] plot [smooth, tension=1] coordinates { (0,0) (-3.5,3.5)};
	
	\draw [line width = .3mm, gray,dashed] plot [smooth, tension=1] (0,7)..controls (-.75,3.5) ..(0,0);
	\draw [line width = .3mm, gray,dashed] plot [smooth, tension=1] (0,7)..controls (-1.6,3.5)..(0,0);
	\draw [line width = .3mm, gray,dashed] plot [smooth, tension=1] (0,7)..controls (-2.4,3.5)..(0,0);
	\draw [line width = .3mm, gray,dashed] plot [smooth, tension=1] (0,7)..controls (-3.3,3.5) ..(0,0);
	
	\draw [line width = .3mm, gray,dashed] plot [smooth, tension=1] (-.9,.9)..controls (-.6,.95) and (-.4,1.1)..(0,1.1);
	\draw [line width = .3mm, gray,dashed] plot [smooth, tension=1] (-2,2)..controls (-1.5,2.2)and (-.8,2.7)..(0,2.8);
	\draw [line width = .3mm, gray,dashed] plot [smooth, tension=1] (-2.8,2.8)..controls (-2,3.3)and (-1,4.4)..(-0,4.5);
	\draw [line width = .3mm, gray,dashed] plot [smooth, tension=1] (-3.4,3.4)..controls  (-3,3.7) and (-1,5.9)..(-0,6);

		\draw[line width = .7mm,red!90!black] (-3.5,3.5)   -- (0,7);
	
	\draw[line width = .5mm] (0,7) -- (0,0);

	\node at (-.9,3.5) {\Huge \RomanNumeralCaps{1}};
	
		\node[rotate = 45] at (-2,5.35) {$\tau =+\infty$};
	\node[rotate = 90] at (.3,3.5) {$\rho =\infty$};
	
	\node[rotate = -45] at (-2,1.6) {$\rho =0$};
	\end{tikzpicture}
	\caption{The $(\tau,\rho)$ coordinate system faithfully covers region \RomanNumeralCaps{1}. Horizontal (vertical) dashed lines are constant $\tau$ (constant $\rho$). The CFT singularity happens infinitely far in the past.}
	\label{compactification2}
\end{figure}

The metric in these coordinates is
\begin{equation}
ds^2 \,=\, d\rho ^2 - 2 d\rho \,d\tau - (e^{2\rho}-1) \,d\tau^2 + e^{2\rho} \,\sum_{i=1}^{d-1} e^{2(p_i-1)\tau} \,dx_i^2\;,
\end{equation}
and the (1+1)-dimensional problem for the field mode $\phi_{\vec{k}}\,(\tau,\rho)$ reduces to 
\begin{equation}
\left(\,(e^{2\rho}-1) \,\partial^2_\rho - 2 \,\partial_\rho \,\partial_\tau + d \,e^{2\rho} \,\partial_\rho - e^{2\rho} \,m^2 - \partial^2_\tau - W_{\vec{k}}\,(\tau)\right) \phi_{\vec{k}}\,(\tau,\rho)\,=\, 0\;, \label{twodimnew}
\end{equation}
with effective potential
\begin{equation}
W_{\vec{k}}\,(\tau) \,=\, \sum_{i=1}^{d-1}\,e^{2\tau (1-p_i)}\;k_i^2  \;.
\end{equation}

\subsection{The zero mode}
\noindent

For $\vec{k} = \vec{0}\;,$ the potential $W_{\vec{k}}(\tau)$ vanishes, and the problem \eqref{twodimnew} becomes time-independent. It is therefore appropriate to use the Fourier basis
\begin{equation}
\phi_{\vec{0},\,\omega }\,(\rho) \,=\,  \int_{-\infty}^{\infty} d\tau \,e^{-i\omega \tau } \;\phi_{\vec{0}}\,(\rho, \tau)\;,
\end{equation}
which has 1-dimensional dynamics 
\begin{equation}
\left( (e^{2\rho}-1) \,\dfrac{d^2}{d \rho^2} +(d e^{2\rho}-2i\omega )\, \dfrac{d}{d \rho}   - e^{2\rho} \,m^2 + \omega^2 \right)  \,\phi_{\vec{0},\,\omega }\,(\rho) = 0 \;,\label{onedimensional}
\end{equation}
with normalizable boundary conditions fixed by the extrapolate dictionary in the new frame
\begin{equation}\label{extrapolate}
\phi_{\vec{0},\,\omega }\,(\rho) \sim e^{-\Delta \rho}\;\mathcal{O}_{\vec{0},\,\omega } \hspace{1.5cm}\text{as}\hspace{.5cm} \rho \rightarrow \infty  \;.
\end{equation}

In order to solve \eqref{onedimensional}, we first rescale the mode $\,\Phi_{\vec{0},\,\omega}\,(\rho) \,=\, e^{i\omega \rho}\,\phi_{\vec{0},\,\omega }\,(\rho)$, and then perform a change of variable $\,\sigma = e^{2\rho}$, to reduce the equation into the hypergeometric equation. It can be solved around the $\rho=\infty$ pole with the correct asymptotic behavior, yielding\footnote{This result is only valid when $\frac{d}{2}-\Delta$ is not an integer. For the integer case, the solution is more involved.}
\begin{equation}
\phi_{\vec{0}}\,(\tau,\rho) \,=\, e^{-\rho\Delta}\,\int_{-\infty}^{\infty} \dfrac{d\omega}{2\pi} \; \mathcal{O}_{\vec{0},\,\omega}\;e^{i\omega \tau }\; \prescript{}{2}{F}_1\left(\dfrac{\Delta-i\omega}{2}\,,\,\dfrac{\Delta-i\omega}{2} \, ,\, 1+ \dfrac{d}{2}-\Delta\,,\, e^{-2\rho}\right)  \;,\label{field12}
\end{equation}
with the hermiticity condition $\,\mathcal{O}_{\vec{0},\,\omega}^\dagger \,=\, \mathcal{O}_{\vec{0},\,-\omega}\;$. These CFT creation/annihilation operators have already been implicitly defined by the extrapolate dictionary \eqref{extrapolate} 
\begin{equation}
\mathcal{O}_{\vec{0},\,\omega}\,=\, \int_{-\infty}^{\infty}d\tau\; e^{-i\omega\tau}\;\mathcal{O}_{\vec{0}}\,(\tau)  \;,
\end{equation}
and plugging this definition in the mode expansion, we get
\begin{gather}
\phi_{\vec{0}}\, (\tau,\rho) \,=\, e^{-\rho\Delta}\;\int_{-\infty}^{\infty} \dfrac{d\omega}{2\pi} \,\int_{-\infty}^{\infty} d\tau' \;e^{-i\omega(\tau'-\tau)}\;\mathcal{O}_{\vec{0}}\,(\tau')\,  \prescript{}{2}{F}_1\left(\dfrac{\Delta-i\omega}{2}\,,\,\dfrac{\Delta-i\omega}{2} \,,\, 1+ \dfrac{d}{2}-\Delta\,,\, e^{-2\rho}\right)=  \nonumber\\ = \int_{-\infty}^{\infty} d\tau' \,K_{\vec{0}}\,(\tau,\rho;\tau')\; \mathcal{O}_{\vec{0}}\,(\tau')  \;,  \label{field}
\end{gather}
where we exchanged the order of integration assuming convergence in the distributional sense, to get that the smearing function is basically a Fourier transform of the hypergeometric function
\begin{equation}
K_{\vec{0}}\,(\tau,\rho;\tau') \,=\, e^{-\rho\Delta}\;\int_{-\infty}^{\infty}\dfrac{d\omega}{2\pi} \;e^{-i\omega(\tau'-\tau)} \; \prescript{}{2}{F}_1\left(\dfrac{\Delta-i\omega}{2}\,,\,\dfrac{\Delta-i\omega}{2} \,,\, 1+ \dfrac{d}{2}-\Delta\,,\, e^{-2\rho}\right) \;. \label{kernel}
\end{equation}

\subsection{Generic modes}
\noindent

Assuming analyticity in $\vec{k}$, we can expand the mode $\phi_{\vec{k}}$ formally as\footnote{The odd terms in the expansion will vanish due to the inversion symmetry.}
\begin{equation}
\phi_{\vec{k}}\, (\tau,\rho)  \,=\, \sum_{n=0}^\infty \,\phi^{(n)}_{\vec{k}}\, (\tau,\rho) \;,
\end{equation}
where $\phi^{(n)}_{\vec{k}}$ is the term of order $ |\vec{k}|^{2n}$ which satisfies the recursive differential equation
\begin{equation}
\left(\,(e^{2\rho}-1) \,\partial^2_\rho - 2 \,\partial_\rho \,\partial_\tau + d \,e^{2\rho} \,\partial_\rho - e^{2\rho} \,m^2 - \partial^2_\tau \right) \phi^{(n)}_{\vec{k}}\,(\tau,\rho)\,=\, W_{\vec{k}}(\tau)\;\phi^{(n-1)}_{\vec{k}}\,(\tau,\rho)\;, \label{twodimnew4}
\end{equation}
for $n\geq 1$, whereas for $\phi^{(0)}_{\vec{k}}$ the equation becomes homogeneous. The extrapolate dictionary will hold independently for each of the $\phi^{(n)}_{\vec{k}}$ since the right hand side of \eqref{twodimnew4} is negligible at the asymptotic boundary
\begin{equation}\label{extrapolate2}
\phi^{(n)}_{\vec{k}}\,(\rho,\tau) \sim e^{-\Delta \rho}\;\mathcal{O}^{(n)}_{\vec{k} }\,(\tau) \hspace{1.5cm}\text{as}\hspace{.5cm} \rho \rightarrow \infty  \;,
\end{equation} 
where the expanded CFT single trace mode $\mathcal{O}^{(n)}_{\vec{k}}\,(\tau)$ corresponds to the order $|\vec{k}|^{2n}$ term in the analogous expansion.

The solution for $n=0$ is obviously the zero mode $\phi_{\vec{k}}^{(0)} \,=\, \phi_{\vec{0}} \, $ and it can be read off from \eqref{field}, with the clear identification $\mathcal{O}^{(0)}_{\vec{k}} \,=\, \mathcal{O}_{\vec{0}}\,$.

For $n=1$, the solution $\phi^{(1)}_{\vec{k}}$ will consist of an homogeneous and an inhomogeneous components. The homogeneous piece will be of the form $\int K_{\vec{k}}^{(0)} \, \mathcal{O}_{\vec{k}}^{(1)}\,$, with $K_{\vec{k}}^{(0)} = K_{\vec{0}}\,$ the one in \eqref{kernel}. The inhomogeneous part will be of the form $\int K_{\vec{k}}^{(1)} \, \mathcal{O}_{\vec{k}}^{(0)}$ with 
\begin{equation}
K^{(1)}_{\vec{k}} = \int_{\text{bulk}} \,G\; W_{\vec{k}}\;  K_{\vec{k}}^{(0)}, \label{two-point}
\end{equation}
where $G$ is a Green's function of the time-independent differential operator on the left hand side of \eqref{twodimnew4}, which can be expanded in terms of hypergeometric functions.

For $n\geq 2$ the only difference is that the inhomogeneous part of $\phi^{(n)}_{\vec{k}}$ will correspond to multiple terms. For instance, for the case $n=2$, there will be two terms $\int K_{\vec{k}}^{(2)} \mathcal{O}^{(0)}_{\vec{k}}\;$ and $ \int K_{\vec{k}}^{(1)} \mathcal{O}^{(1)}_{\vec{k}}\; $, where $K_{\vec{k}}^{(2)}$ corresponds to two bulk vertices of the form \eqref{two-point}. 

The complete formal resummation yields
\begin{equation}
\phi_{\vec{k}} \,= \, \int K_{\vec{k}}^{(0)}\, \mathcal{O}_{\vec{k}}^{(0)} \,+\, \int K_{\vec{k}}^{(1)}\, \mathcal{O}_{\vec{k}}^{(0)} \,+\, \int K_{\vec{k}}^{(0)}\, \mathcal{O}_{\vec{k}}^{(1)} \,+\, ...\, = \,\int K_{\vec{k}}\;\mathcal{O}_{\vec{k}}
\end{equation}

\vspace{.25cm}

\bibliographystyle{style}
\bibliography{Bibliography2.bib}

\providecommand{\href}[2]{#2}\begingroup\raggedright\begin{thebibliography}{10}

\bibitem{Banks:1998dd}
T.~Banks, M.~R. Douglas, G.~T. Horowitz, and E.~J. Martinec, ``{AdS dynamics
  from conformal field theory},''
\href{http://arxiv.org/abs/hep-th/9808016}{{\ttfamily arXiv:hep-th/9808016
  [hep-th]}}.

\bibitem{Bena:1999jv}
I.~Bena, ``{On the construction of local fields in the bulk of AdS(5) and other
  spaces},'' \href{http://dx.doi.org/10.1103/PhysRevD.62.066007}{{\em Phys.
  Rev.} {\bfseries D62} (2000) 066007},
\href{http://arxiv.org/abs/hep-th/9905186}{{\ttfamily arXiv:hep-th/9905186
  [hep-th]}}.

\bibitem{Hamilton:2005ju}
A.~Hamilton, D.~N. Kabat, G.~Lifschytz, and D.~A. Lowe, ``{Local bulk operators
  in AdS/CFT: A Boundary view of horizons and locality},''
  \href{http://dx.doi.org/10.1103/PhysRevD.73.086003}{{\em Phys. Rev.}
  {\bfseries D73} (2006) 086003},
\href{http://arxiv.org/abs/hep-th/0506118}{{\ttfamily arXiv:hep-th/0506118
  [hep-th]}}.

\bibitem{Hamilton:2006az}
A.~Hamilton, D.~N. Kabat, G.~Lifschytz, and D.~A. Lowe, ``{Holographic
  representation of local bulk operators},''
  \href{http://dx.doi.org/10.1103/PhysRevD.74.066009}{{\em Phys. Rev.}
  {\bfseries D74} (2006) 066009},
\href{http://arxiv.org/abs/hep-th/0606141}{{\ttfamily arXiv:hep-th/0606141
  [hep-th]}}.

\bibitem{Kabat:2011rz}
D.~Kabat, G.~Lifschytz, and D.~A. Lowe, ``{Constructing local bulk observables
  in interacting AdS/CFT},''
  \href{http://dx.doi.org/10.1103/PhysRevD.83.106009}{{\em Phys. Rev.}
  {\bfseries D83} (2011) 106009},
\href{http://arxiv.org/abs/1102.2910}{{\ttfamily arXiv:1102.2910 [hep-th]}}.

\bibitem{Heemskerk:2012mn}
I.~Heemskerk, D.~Marolf, J.~Polchinski, and J.~Sully, ``{Bulk and Transhorizon
  Measurements in AdS/CFT},''
  \href{http://dx.doi.org/10.1007/JHEP10(2012)165}{{\em JHEP} {\bfseries 10}
  (2012) 165},
\href{http://arxiv.org/abs/1201.3664}{{\ttfamily arXiv:1201.3664 [hep-th]}}.

\bibitem{Kabat:2012av}
D.~Kabat and G.~Lifschytz, ``{CFT representation of interacting bulk gauge
  fields in AdS},'' \href{http://dx.doi.org/10.1103/PhysRevD.87.086004}{{\em
  Phys. Rev.} {\bfseries D87} no.~8, (2013) 086004},
\href{http://arxiv.org/abs/1212.3788}{{\ttfamily arXiv:1212.3788 [hep-th]}}.

\bibitem{Kabat:2012hp}
D.~Kabat, G.~Lifschytz, S.~Roy, and D.~Sarkar, ``{Holographic representation of
  bulk fields with spin in AdS/CFT},''
  \href{http://dx.doi.org/10.1103/PhysRevD.86.026004,
  10.1103/PhysRevD.86.029901}{{\em Phys. Rev.} {\bfseries D86} (2012) 026004},
\href{http://arxiv.org/abs/1204.0126}{{\ttfamily arXiv:1204.0126 [hep-th]}}.

\bibitem{Kabat:2013wga}
D.~Kabat and G.~Lifschytz, ``{Decoding the hologram: Scalar fields interacting
  with gravity},'' \href{http://dx.doi.org/10.1103/PhysRevD.89.066010}{{\em
  Phys. Rev.} {\bfseries D89} no.~6, (2014) 066010},
\href{http://arxiv.org/abs/1311.3020}{{\ttfamily arXiv:1311.3020 [hep-th]}}.

\bibitem{Jafferis:2015del}
D.~L. Jafferis, A.~Lewkowycz, J.~Maldacena, and S.~J. Suh, ``{Relative entropy
  equals bulk relative entropy},''
  \href{http://dx.doi.org/10.1007/JHEP06(2016)004}{{\em JHEP} {\bfseries 06}
  (2016) 004},
\href{http://arxiv.org/abs/1512.06431}{{\ttfamily arXiv:1512.06431 [hep-th]}}.

\bibitem{Jafferis:2014lza}
D.~L. Jafferis and S.~J. Suh, ``{The Gravity Duals of Modular Hamiltonians},''
  \href{http://dx.doi.org/10.1007/JHEP09(2016)068}{{\em JHEP} {\bfseries 09}
  (2016) 068},
\href{http://arxiv.org/abs/1412.8465}{{\ttfamily arXiv:1412.8465 [hep-th]}}.

\bibitem{Dong:2016eik}
X.~Dong, D.~Harlow, and A.~C. Wall, ``{Reconstruction of Bulk Operators within
  the Entanglement Wedge in Gauge-Gravity Duality},''
  \href{http://dx.doi.org/10.1103/PhysRevLett.117.021601}{{\em Phys. Rev.
  Lett.} {\bfseries 117} no.~2, (2016) 021601},
\href{http://arxiv.org/abs/1601.05416}{{\ttfamily arXiv:1601.05416 [hep-th]}}.

\bibitem{Papadodimas:2012aq}
K.~Papadodimas and S.~Raju, ``{An Infalling Observer in AdS/CFT},''
  \href{http://dx.doi.org/10.1007/JHEP10(2013)212}{{\em JHEP} {\bfseries 10}
  (2013) 212},
\href{http://arxiv.org/abs/1211.6767}{{\ttfamily arXiv:1211.6767 [hep-th]}}.

\bibitem{Papadodimas:2013jku}
K.~Papadodimas and S.~Raju, ``{State-Dependent Bulk-Boundary Maps and Black
  Hole Complementarity},''
  \href{http://dx.doi.org/10.1103/PhysRevD.89.086010}{{\em Phys. Rev.}
  {\bfseries D89} no.~8, (2014) 086010},
\href{http://arxiv.org/abs/1310.6335}{{\ttfamily arXiv:1310.6335 [hep-th]}}.

\bibitem{deBoer:2018ibj}
J.~de~Boer, R.~Van~Breukelen, S.~F. Lokhande, K.~Papadodimas, and E.~Verlinde,
  ``{On the interior geometry of a typical black hole microstate},''
  \href{http://dx.doi.org/10.1007/JHEP05(2019)010}{{\em JHEP} {\bfseries 05}
  (2019) 010},
\href{http://arxiv.org/abs/1804.10580}{{\ttfamily arXiv:1804.10580 [hep-th]}}.

\bibitem{deBoer:2019kyr}
J.~De~Boer, R.~Van~Breukelen, S.~F. Lokhande, K.~Papadodimas, and E.~Verlinde,
  ``{Probing typical black hole microstates},''
\href{http://arxiv.org/abs/1901.08527}{{\ttfamily arXiv:1901.08527 [hep-th]}}.

\bibitem{Faulkner:2017vdd}
T.~Faulkner and A.~Lewkowycz, ``{Bulk locality from modular flow},''
  \href{http://dx.doi.org/10.1007/JHEP07(2017)151}{{\em JHEP} {\bfseries 07}
  (2017) 151},
\href{http://arxiv.org/abs/1704.05464}{{\ttfamily arXiv:1704.05464 [hep-th]}}.

\bibitem{Engelhardt:2013jda}
N.~Engelhardt and G.~T. Horowitz, ``{Entanglement Entropy Near Cosmological
  Singularities},'' \href{http://dx.doi.org/10.1007/JHEP06(2013)041}{{\em JHEP}
  {\bfseries 06} (2013) 041},
\href{http://arxiv.org/abs/1303.4442}{{\ttfamily arXiv:1303.4442 [hep-th]}}.

\bibitem{Engelhardt:2014mea}
N.~Engelhardt, T.~Hertog, and G.~T. Horowitz, ``{Holographic Signatures of
  Cosmological Singularities},''
  \href{http://dx.doi.org/10.1103/PhysRevLett.113.121602}{{\em Phys. Rev.
  Lett.} {\bfseries 113} (2014) 121602},
\href{http://arxiv.org/abs/1404.2309}{{\ttfamily arXiv:1404.2309 [hep-th]}}.

\bibitem{Engelhardt:2015gta}
N.~Engelhardt, T.~Hertog, and G.~T. Horowitz, ``{Further Holographic
  Investigations of Big Bang Singularities},''
  \href{http://dx.doi.org/10.1007/JHEP07(2015)044}{{\em JHEP} {\bfseries 07}
  (2015) 044},
\href{http://arxiv.org/abs/1503.08838}{{\ttfamily arXiv:1503.08838 [hep-th]}}.

\bibitem{Barbon:2015ria}
J.~L.~F. Barbon and E.~Rabinovici, ``{Holographic complexity and spacetime
  singularities},'' \href{http://dx.doi.org/10.1007/JHEP01(2016)084}{{\em JHEP}
  {\bfseries 01} (2016) 084},
\href{http://arxiv.org/abs/1509.09291}{{\ttfamily arXiv:1509.09291 [hep-th]}}.

\bibitem{Hubeny:2012wa}
V.~E. Hubeny and M.~Rangamani, ``{Causal Holographic Information},''
  \href{http://dx.doi.org/10.1007/JHEP06(2012)114}{{\em JHEP} {\bfseries 06}
  (2012) 114},
\href{http://arxiv.org/abs/1204.1698}{{\ttfamily arXiv:1204.1698 [hep-th]}}.

\bibitem{Wall:2012uf}
A.~C. Wall, ``{Maximin Surfaces, and the Strong Subadditivity of the Covariant
  Holographic Entanglement Entropy},''
  \href{http://dx.doi.org/10.1088/0264-9381/31/22/225007}{{\em Class. Quant.
  Grav.} {\bfseries 31} no.~22, (2014) 225007},
\href{http://arxiv.org/abs/1211.3494}{{\ttfamily arXiv:1211.3494 [hep-th]}}.

\bibitem{Headrick:2014cta}
M.~Headrick, V.~E. Hubeny, A.~Lawrence, and M.~Rangamani, ``{Causality \&
  holographic entanglement entropy},''
  \href{http://dx.doi.org/10.1007/JHEP12(2014)162}{{\em JHEP} {\bfseries 12}
  (2014) 162},
\href{http://arxiv.org/abs/1408.6300}{{\ttfamily arXiv:1408.6300 [hep-th]}}.

\bibitem{Maldacena:2010un}
J.~Maldacena, ``{Vacuum decay into Anti de Sitter space},''
\href{http://arxiv.org/abs/1012.0274}{{\ttfamily arXiv:1012.0274 [hep-th]}}.

\bibitem{Harlow:2010my}
D.~Harlow and L.~Susskind, ``{Crunches, Hats, and a Conjecture},''
\href{http://arxiv.org/abs/1012.5302}{{\ttfamily arXiv:1012.5302 [hep-th]}}.

\bibitem{Barbon:2011ta}
J.~L.~F. Barbon and E.~Rabinovici, ``{AdS Crunches, CFT Falls And Cosmological
  Complementarity},'' \href{http://dx.doi.org/10.1007/JHEP04(2011)044}{{\em
  JHEP} {\bfseries 04} (2011) 044},
\href{http://arxiv.org/abs/1102.3015}{{\ttfamily arXiv:1102.3015 [hep-th]}}.

\end{thebibliography}\endgroup

\end{document}